\documentclass[reprint, notitlepage,superscriptaddress,preprintnumbers,nofootinbib,nobibnotes,amsmath,amssymb,aps,twocolumn]{revtex4-1}
\usepackage{graphicx,grffile}
\usepackage{dcolumn}
\usepackage{bm}
\usepackage{hyperref}
\usepackage{aas_macros,braket}

\begin{document}

\title{Polypolar spherical harmonic decomposition of galaxy correlators in redshift space: Toward testing cosmic rotational symmetry}
\author{Maresuke Shiraishi}
\affiliation{%
  Kavli Institute for the Physics and Mathematics of the Universe (Kavli IPMU, WPI), UTIAS, The University of Tokyo, Chiba, 277-8583, Japan
}
\author{Naonori S. Sugiyama}
\affiliation{%
  Kavli Institute for the Physics and Mathematics of the Universe (Kavli IPMU, WPI), UTIAS, The University of Tokyo, Chiba, 277-8583, Japan
}
\author{Teppei Okumura}
\affiliation{%
  Kavli Institute for the Physics and Mathematics of the Universe (Kavli IPMU, WPI), UTIAS, The University of Tokyo, Chiba, 277-8583, Japan
}
\affiliation{%
  Institute of Astronomy and Astrophysics, Academia Sinica, P. O. Box 23-141, Taipei 10617, Taiwan
  }

\date{\today}

\begin{abstract}
  We propose an efficient way to test rotational invariance in the cosmological perturbations by use of galaxy correlation functions. In symmetry-breaking cases, the galaxy power spectrum can have extra angular dependence in addition to the usual one due to the redshift-space distortion, $ \hat{k} \cdot \hat{n}$. We confirm that, via the decomposition into not the usual Legendre basis ${\cal  L}_\ell(\hat{k} \cdot \hat{n})$ but the bipolar spherical harmonic one $\{Y_{\ell}(\hat{k}) \otimes Y_{\ell'}(\hat{n})\}_{LM}$, the symmetry-breaking signal can be completely distinguished from the usual isotropic one since the former yields nonvanishing $L \geq 1$ modes but the latter is confined to the $L = 0$ one. As a demonstration, we analyze the signatures due to primordial-origin symmetry breakings such as the well-known quadrupolar-type and dipolar-type power asymmetries and find nonzero $L = 2$ and $1$ modes, respectively. Fisher matrix forecasts of their constraints indicate that the {\it Planck}-level sensitivity could be achieved by the SDSS or BOSS-CMASS data, and an order-of-magnitude improvement is expected in a near future survey as PFS or Euclid by virtue of an increase in accessible Fourier mode. Our methodology is model-independent and hence applicable to the searches for various types of statistically anisotropic fluctuations. 
\end{abstract} 

%\pacs{98.80.-k}

\maketitle

%%%%%%%%%%%%%%%%%%%%%%%%%%%%%%%%%%%%%%%%%%%%%%%%%%%%%%%%%%%%%%%%%%%%%%%%%%%%%
\section{Introduction}
%%%%%%%%%%%%%%%%%%%%%%%%%%%%%%%%%%%%%%%%%%%%%%%%%%%%%%%%%%%%%%%%%%%%%%%%%%%%%

Symmetries give a basic guideline for building a cosmological model, and determine the statistical property of the resulting cosmological perturbations. This paper focuses on isotropy (rotational symmetry) as a key observational indicator in cosmology. A concordance model of cosmology, such as a $\Lambda$CDM model or a single-field slow-roll inflation model, predicts nearly isotropic and homogeneous cosmological fluctuations. The observed cosmic microwave background (CMB) anisotropies indicate the smallness of the symmetry breakings \cite{Kim:2013gka,Ade:2015lrj,Ade:2015hxq,Aiola:2015rqa,Saadeh:2016sak}, supporting such a concordance scenario, however, there are still room for many alternatives (e.g. \cite{Erickcek:2008sm,Erickcek:2009at,Pontzen:2010eg,Dimastrogiovanni:2010sm,Soda:2012zm,Maleknejad:2012fw,Bartolo:2012sd,Naruko:2014bxa,Bartolo:2014hwa,Bartolo:2015dga,Bartolo:2013msa, Bartolo:2014xfa,Jazayeri:2014nya,Firouzjahi:2016fxf}).

For more stringent tests of these models, multidirectional analyses with other observables are indispensable. There are already many phenomenological studies on e.g. galaxies \cite{Ando:2008zza,Pullen:2010zy,Jeong:2012df,Dai:2013kra,Baghram:2013lxa,Dai:2015wla,Emami:2015uva,Raccanelli:2015oma}, gravitational lensing \cite{Hassani:2015zat,Pitrou:2015iya,Pereira:2015jya}, 21-cm fluctuations \cite{Shiraishi:2016omb} and CMB spectral distortions \cite{Shiraishi:2015lma,Shiraishi:2016hjd}. These observables can yield the constraints at scales and redshifts unaccessible by the analysis of the CMB anisotropies, which most of the preceding studies focused on \cite{Hirata:2009ar,Pullen:2010zy,Rubart:2013tx,Kothari:2013gya,Alonso:2014xca,Eggemeier:2015ifa,Bengaly:2015xkw,Bengaly:2016amk,Javanmardi:2016whx}. 

As an observable beyond the CMB anisotropies, we here study the 2-point correlation function of galaxy distributions. In Ref.~\cite{Pullen:2010zy}, the constraint on a statistically anisotropic model from the SDSS data was obtained by employing the angular correlation function defined in 2D harmonic space. On the other hand, there has been no analysis of cosmic statistical anisotropy based on the galaxy clustering in full 3D space because such an analysis becomes much more complicated than the 2D case. However, the cosmological information in the 3D clustering is expected to be much larger since the number of Fourier modes is proportional to $k^3$. This fact motivates us to develop an accurate method in 3D because ongoing and future galaxy surveys can probe larger and larger volume.

The main goal of this paper is therefore to find an efficient way to extract the anisotropic information of any cosmological source by use of the 3D galaxy correlation function. In usual isotropic and homogeneous cases, the angular dependence in the observed galaxy power spectrum is quantified by the radial components of peculiar velocities of galaxies, known as redshift-space distortions, and thus can be completely decomposed using the Legendre polynomials ${\cal L}_\ell(\hat{k} \cdot \hat{n})$, where $\hat{k} \cdot \hat{n}$ is the cosine of the angle between a wave vector and a line-of-sight direction. In our cases, however, the breaking of isotropy induces extra angular dependences and the Legendre expansion would fail to capture the full angular dependences. We hence consider the decomposition into the bipolar spherical harmonic (BipoSH) basis $\{Y_{\ell}(\hat{k}) \otimes Y_{\ell'}(\hat{n})\}_{LM}$ \cite{Varshalovich:1988ye}. This kind of basis was previously utilized to deal with the CMB statistical anisotropy \cite{Hajian:2003qq,Hajian:2005jh,Basak:2006ew,Pullen:2007tu,Book:2011na} and the wide-angle effect in the galaxy survey \cite{Heavens:1994iq,Hamilton:1995px,Szalay:1997cc, Szapudi:2004gh,Papai:2008bd,Bertacca:2012tp,Raccanelli:2013dza}. In the main text of this paper, we discuss the symmetry-breaking signatures by means of the formalism not including the wide-angle effect. We then find that, by virtue of the BipoSH decomposition, the signal due to anisotropy can be efficiently filtered from the galaxy power spectrum because of the presence of nonzero $L \geq 1$ modes.%
\footnote{
As shown in Appendix~\ref{appen:Bl1l2Lx}, even if the wide-angle effect exists, the statistically-anisotropic signatures are completely distinguishable according to the tripolar spherical harmonic (TripoSH) decomposition \cite{Varshalovich:1988ye}.
}

After studying the formalism, we demonstrate its usage by analyzing two specific models of the anisotropic galaxy power spectra. The first model contains the quadrupolar directional dependence, $(\hat{ k} \cdot \hat{p})^2$, where $\hat{p}$ is some preferred direction. This can be realized e.g. in the inflationary models involving the vector field \cite{Dimastrogiovanni:2010sm,Soda:2012zm,Maleknejad:2012fw,Bartolo:2012sd,Naruko:2014bxa,Bartolo:2014hwa,Bartolo:2015dga}, or from an inflating solid or elastic medium \cite{Bartolo:2013msa, Bartolo:2014xfa}. These models also predict symmetry-breaking non-Gaussianities \cite{Bartolo:2012sd,Bartolo:2013msa,Shiraishi:2013vja,Shiraishi:2013oqa,Abolhasani:2013zya,Bartolo:2015dga,Shiraishi:2016mok}. The second one contains the dipolar modulation, $\hat{x} \cdot \hat{p}$, which is detected from large-scale CMB data \cite{Ade:2015lrj,Ade:2015hxq,Aiola:2015rqa}. Such term may be related to e.g. supercurvature fluctuations in the inflationary era \cite{Erickcek:2008sm,Erickcek:2009at,Kanno:2013ohv,Byrnes:2016uqw} or primordial non-Gaussianities \cite{Schmidt:2012ky,Lyth:2013vha, Abolhasani:2013vaa,Ashoorioon:2015pia,Ashoorioon:2016lrg}. Strictly speaking, this type of modulation also breaks statistical homogeneity. However, the deviation from homogeneity is negligibly small; thus, one may assume the translation invariance in the phenomenological study. We show that the BipoSH decomposition successfully extracts the distinctive signal coming from these directional-dependent terms because of nonzero $L = 2$ and $1$ modes. The detectability of the symmetry-breaking signal generated in these models is estimated via the Fisher matrix computations. In this paper we consider four generation surveys: the Sloan Digital Sky Survey (SDSS) \cite{Abazajian:2008wr}, the Baryon Oscillation Spectroscopic Survey (BOSS) \cite{Bolton:2012hz,Dawson:2012va} that is part of SDSS-III \cite{Eisenstein:2011sa}, the Subaru Prime Focus Spectrograph (PFS) \cite{Ellis:2012rn}, and Euclid \cite{Laureijs:2011gra}. We then find that the analysis with the BipoSH coefficients could realize the sensitivity comparable to (beyond) the {\it Planck} results \cite{Ade:2015lrj,Ade:2015hxq,Aiola:2015rqa} in a current (futuristic) galaxy survey. 

This paper is organized as follows. In the next section, we introduce the BipoSH decomposition to extract the anisotropic signal from the 3D galaxy correlation function, and check the response of the BipoSH coefficients to primordial-origin quadrupolar and dipolar power asymmetries. In Sec.~\ref{sec:Fish}, we compute the Fisher matrix and find minimum detectable amplitudes of the quadrupolar and dipolar asymmetries in the past, present and futuristic galaxy surveys. Section~\ref{sec:conclusion} concludes this paper. In Appendix~\ref{appen:Cl}, we explain how to estimate minimum detectable amplitudes of the quadrupolar and dipolar asymmetries from the 2D angular correlation, which are compared with the 3D results in Sec.~\ref{sec:Fish}. In Appendix~\ref{appen:Bl1l2Lx}, we summarize a complete decomposition technique using the TripoSH basis for the 3D galaxy correlation function including the wide-angle effect. Mathematical identities used for derivations are summarized in Appendix~\ref{appen:math}.

%%%%%%%%%%%%%%%%%%%%%%%%%%%%%%%%%%%%%%%%%%%%%%%%%%%%%%%%%%%%%%%%%%%%%%%%%%%%%
\section{BipoSH decomposition of galaxy correlation functions} \label{sec:formalism} 
%%%%%%%%%%%%%%%%%%%%%%%%%%%%%%%%%%%%%%%%%%%%%%%%%%%%%%%%%%%%%%%%%%%%%%%%%%%%%

Direct observables in galaxy surveys are the 3D positions of galaxies in redshift space, ${\bf x}$.
Thus let us begin with the redshift-space overdensity of galaxies, $\delta^s({\bf x},z)$, where the superscript $s$ denotes a quantity defined in redshift space. An argument of time, namely redshift $z$, will be hereinafter omitted in some variables for simplicity. The 2-point galaxy correlation function $\xi^s \equiv \Braket{\delta^s({\bf x}_1) \delta^s({\bf x}_2)} $ is characterized by the three directions, $\hat{x}_1$, $\hat{x}_2$ and $\hat{x}_{12} \equiv \widehat{{\bf x}_1 - {\bf x}_2}$, and hence decomposed using the TripoSH basis $\{ Y_{\ell}(\widehat{x}_{12}) \otimes \{ Y_{\ell_1}(\hat{x}_1) \otimes Y_{\ell_2}(\hat{x}_2) \}_{\ell'} \}_{LM} $ \cite{Varshalovich:1988ye,Szalay:1997cc,Szapudi:2004gh,Papai:2008bd,Bertacca:2012tp,Raccanelli:2013dza}. This is a complete treatment but not suitable for a practical analysis because of the computational complexity. We therefore adopt the so-called local plane parallel approximation $\hat{x}_1 = \hat{x}_2 \equiv \hat{n}$ in the following discussions (see Appendix~\ref{appen:Bl1l2Lx} for a complete analysis without this approximation). This approximation is justified as long as the visual angle for the correlation scales of interest ${\bf x}_{12}$ is small. Under this approximation, the 2-point correlation function can be expanded according to
\begin{eqnarray}
  \xi^s({\bf x}_{12}, \hat{n}) &=& \sum_{\ell \ell'  L M } \xi_{\ell \ell'}^{LM}(x_{12}) X_{\ell \ell'}^{LM}(\hat{x}_{12},\hat{n}) ~, 
\end{eqnarray}
where the BipoSH basis \cite{Varshalovich:1988ye} reads
\begin{eqnarray}
  X_{\ell \ell'}^{LM}(\hat{x}_{12},\hat{n})
  &\equiv& \{Y_{\ell}(\hat{x}_{12}) \otimes Y_{\ell'}(\hat{n})\}_{LM} \nonumber \\ 
  &=& 
  \sum_{mm'} {\cal C}_{\ell m \ell' m'}^{LM} Y_{\ell m}(\hat{x}_{12}) Y_{\ell' m'}(\hat{n})  \label{eq:S_basis}
\end{eqnarray}
with ${\cal C}_{l_1 m_1 l_2 m_2}^{l_3 m_3} \equiv (-1)^{l_1 - l_2 + m_3} \sqrt{2l_3 + 1} \left( \begin{smallmatrix} l_1 & l_2 & l_3 \\ m_1 & m_2 & -m_3 \end{smallmatrix}  \right)$ denoting the Clebsch-Gordan coefficients. The orthonormality of the BipoSH,
\begin{eqnarray}
 && \int d^2 \hat{x}_{12} \int d^2 \hat{n} X_{\ell \ell'}^{L M}(\hat{x}_{12},\hat{n}) X_{\tilde{\ell} \tilde{\ell}'}^{\tilde{L} \tilde{M} *}(\hat{x}_{12},\hat{n}) \nonumber \\ 
  &&\qquad = \delta_{L, \tilde{L}} \delta_{M, \tilde{M}}   
  \delta_{\ell, \tilde{\ell}} \delta_{\ell', \tilde{\ell}'} ~,
\end{eqnarray}
yields the translation law
\begin{eqnarray}
  \xi_{\ell\ell'}^{LM}(x_{12}) 
  = \int d^2 \hat{x}_{12} \int d^2 \hat{n} \xi^s({\bf x}_{12}, \hat{n})
  X_{\ell \ell'}^{LM *}(\hat{x}_{12},\hat{n}) .
  \end{eqnarray}
An interesting property of this BipoSH decomposition is that the isotropic information is completely confined to the zero total angular momentum modes $\xi_{\ell \ell'}^{00}(x_{12})$. In other words, breaking rotational symmetry is required for the generation of nonvanishing $L \geq 1$ modes and therefore they will become clean observables of cosmic statistical anisotropy. The identical property is also seen in more general expression without the local plane parallel approximation as shown in Appendix~\ref{appen:Bl1l2Lx}

To investigate the signal expected from the theoretical models, we move to the Fourier space, according to
\begin{eqnarray}
  \delta^s({\bf x})
  &=&  \int \frac{d^3 k}{(2\pi)^3} \delta^s({\bf k}) 
 e^{i {\bf k} \cdot {\bf x}} ~. \label{eq:deltag}
\end{eqnarray}
In the models focused on below, the departure from statistical homogeneity in the primordial power spectrum is completely absent or negligibly small; thus, the galaxy power spectrum is always written as
\begin{eqnarray}
  \Braket{\delta^s({\bf k}_1) \delta^s({\bf k}_2) }
  = (2\pi)^3 P^s({\bf k}_1, \hat{n}) \delta^{(3)}\left({\bf k}_1 + {\bf k}_2 \right) ~.
\end{eqnarray}
When this is decomposed by following
\begin{eqnarray}
 P^s({\bf k}, \hat{n}) = \sum_{\ell \ell' LM} \pi_{\ell \ell'}^{LM}(k) X_{\ell \ell'}^{LM}(\hat{k},\hat{n}) ~,
\end{eqnarray}
the expansion coefficients are related to the real-space ones according to the Hankel transformation:
\begin{eqnarray}
    \xi_{\ell\ell'}^{LM}(x_{12}) = 
    i^{\ell} \int_0^\infty \frac{k^2 d k}{2\pi^2} j_{\ell}(k x_{12}) \pi_{\ell \ell'}^{LM}(k) ~. 
\end{eqnarray}
To derive this, we have used Eqs.~\eqref{eq:e_expand} and \eqref{eq:3j_sum_m1m2}. The translation law reads 
\begin{eqnarray}
  \pi_{\ell\ell'}^{LM}(k) &=&  \int d^2 \hat{k} \int d^2 \hat{n} P^s({\bf k}, \hat{n}) X_{\ell \ell'}^{LM *}(\hat{k},\hat{n})  ~. \label{eq:PlldLM_def}
\end{eqnarray}

In the following, we assume that the Universe is rotationally asymmetric during inflation, but after that, it is completely isotropized and the density fluctuations grow linearly. Then Eq.~(\ref{eq:deltag}) is simply expressed as \cite{Kaiser:1987qv,Hamilton:1997zq}
\begin{eqnarray}
 \delta^s({\bf x})
  = \int \frac{d^3 k}{(2\pi)^3} \delta_{m}({\bf k}) \left[b + f (\hat{k} \cdot \hat{x})^2 \right] e^{i {\bf k} \cdot {\bf x}} ~,
\label{eq:deltag_linear}
\end{eqnarray}
where $b(z)$ is a scale-independent bias parameter. The matter fluctuation $\delta_m$ is linearly related to the primordial curvature perturbation $\zeta$ according to $\delta_m({\bf k}, z) = M_k(z) \zeta_{\bf k}$, with $M_k(z)$ being the matter transfer function. The prefactor for the redshift-space distortion, $f(z)$, is a function of the growth factor $D(a)$, reading $f \equiv \partial \ln D / \partial \ln a$ with $a$ denoting the scale factor. For simplicity, let us focus on the equal-time ($z_1 = z_2$) correlation alone. In the local plane-parallel limit, the galaxy power spectrum reads
\begin{eqnarray}
  P^s({\bf k}, \hat{n}) = P_m({\bf k})
  \left[b + f (\hat{k} \cdot \hat{n})^2 \right]^2 ~, \label{eq:deltag2}
\end{eqnarray}
where $P_m$ is the dark matter power spectrum in real space.

In the case that $P_m$ is independent of $\hat{k}$ and $\hat{n}$; namely, the Universe is isotropic, via the computations described in Appendix~\ref{appen:math}, one can obtain
\begin{eqnarray}
  \pi_{\ell \ell'}^{LM}(k) &=& 
P_{\ell}(k) \frac{4\pi }{ 2\ell+1 } H_{\ell \ell 0}^{-1} 
\delta_{\ell, \ell'} \delta_{L,0}\delta_{M,0} ~, \label{eq:PlldLM_iso}
\end{eqnarray}
where $H_{l_1 l_2 l_3} \equiv \left( \begin{smallmatrix} l_1 & l_2 & l_3 \\ 0 & 0 & 0 \end{smallmatrix}  \right)$ and 
\begin{eqnarray}
 P_0(k) &=& \left(b^2 + \frac{2}{3}b f + \frac{1}{5} f^2 \right) M_k^2 P_\zeta(k)   ~, \\
 P_2(k)  &=& \left( \frac{4}{3}b f  + \frac{4}{7} f^2 \right) M_k^2 P_\zeta(k)   ~, \\
 P_4(k)   &=& \frac{8}{35} f^2 M_k^2 P_\zeta(k)  ~, \\
 P_1(k) &=& P_3(k) = P_{\ell \geq 5}(k) = 0 ~,
\end{eqnarray}
with $P_\zeta$ being the isotropic primordial curvature power spectrum. As clearly seen here, the signal is confined to $L = 0$. In contrast, as demonstrated below, the breaking of rotational symmetry in $P_m$ can induce nonvanishing signal for $L \geq 1$. We here demonstrate it by investigating the signal generated in two popular primordial-origin symmetry-breaking models, i.e., the so-called quadrupolar and dipolar rotational asymmetries. 

For such analyses, let us introduce the {\em reduced} coefficients 
\begin{equation}
  P_{\ell \ell'}^{L M} \equiv 
  \pi_{\ell \ell'}^{LM} (-1)^L  
  \sqrt{\frac{(2L+1)(2\ell+1)(2\ell'+1)}{(4\pi)^2 } } H_{\ell \ell' L}
  \label{eq:pi2P},
\end{equation}
where $H_{\ell\ell'L}$ filters even~$\ell + \ell' + L$ components. This normalizes the BipoSH coefficients as $P_{\ell\ell}^{00}(k)$ recovers the usual Legendre coefficients $P_\ell(k)$ \cite{Hamilton:1997zq} [see Eq.~\eqref{eq:PlldLM_iso}]. In the models discussed below, $P_{\ell \ell'}^{L M}(k)$ has the equal information to $\pi_{\ell \ell'}^{LM}(k)$ because of the absence of the odd $\ell + \ell' + L$ signal in $\pi_{\ell \ell'}^{LM}(k)$.

We begin from the analysis of the quadrupolar asymmetry model.

\begin{figure*}[t]
  \begin{tabular}{cc} 
   \begin{minipage}{0.5\hsize}
     \begin{center}
 \includegraphics[width=1\textwidth]{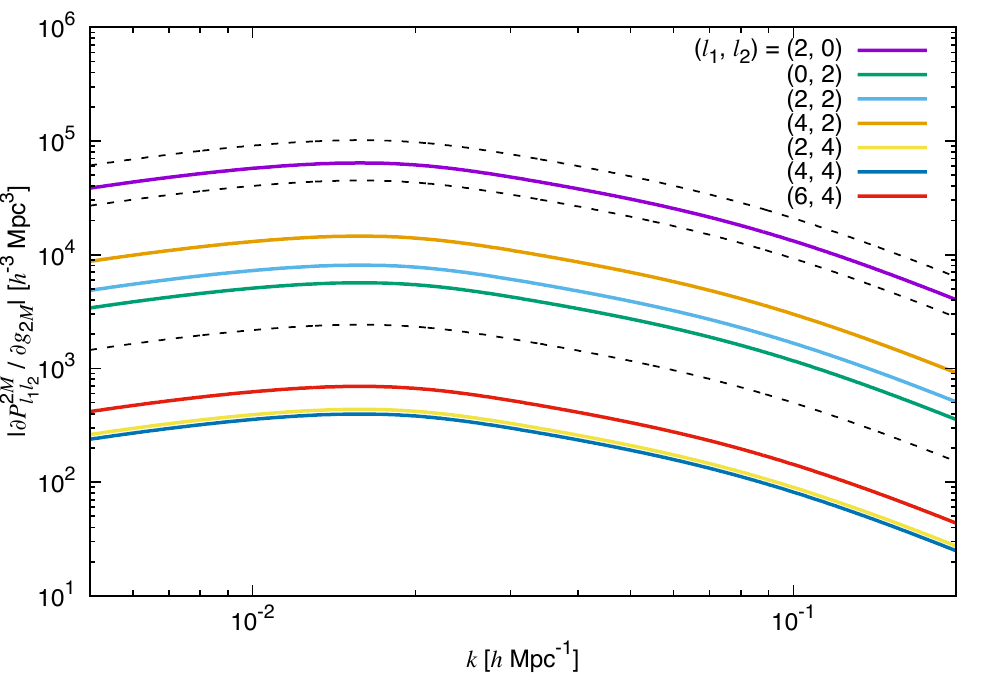}
  \end{center}
   \end{minipage}
   \begin{minipage}{0.5\hsize}
     \begin{center}
 \includegraphics[width=1\textwidth]{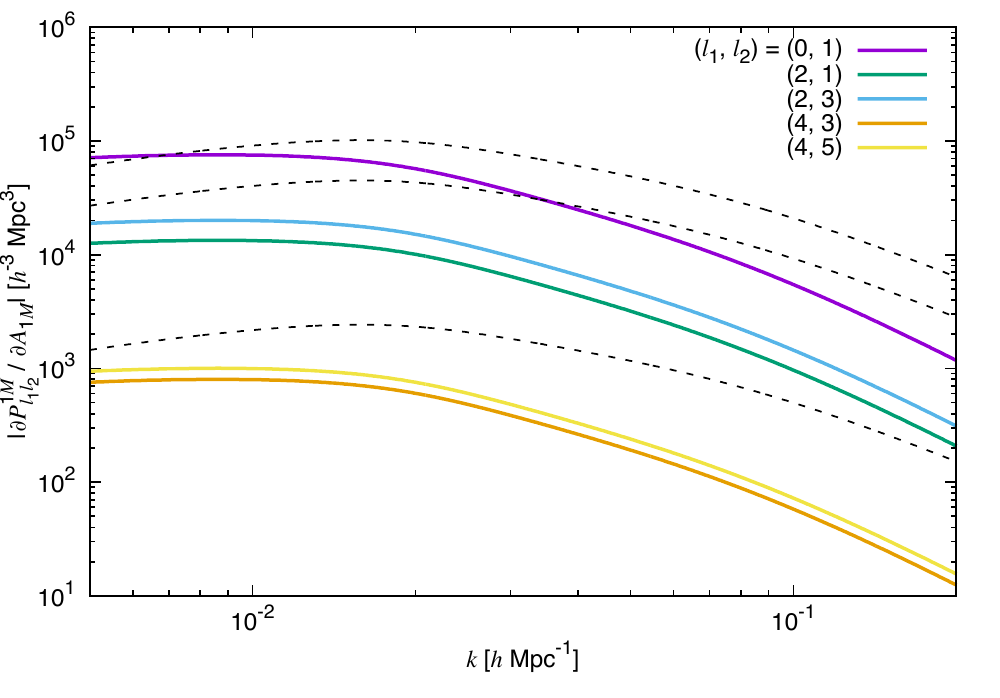}
  \end{center}
   \end{minipage}
 \end{tabular}
  \\
 \begin{tabular}{cc}
   \begin{minipage}{0.5\hsize}
     \begin{center}
   \includegraphics[width=1\textwidth]{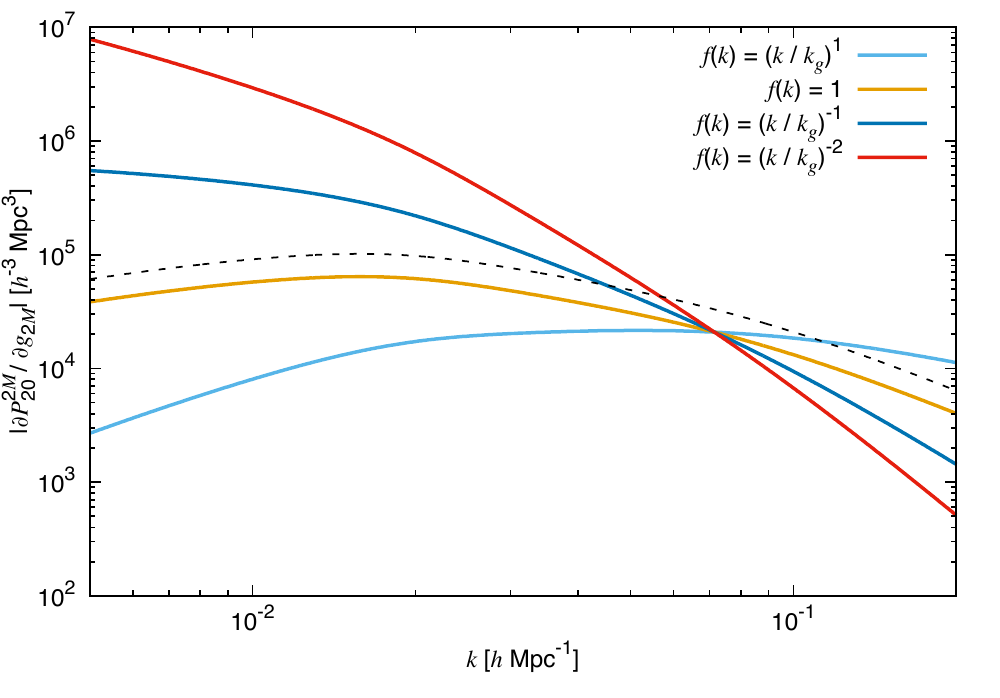}
  \end{center}
   \end{minipage}
   \begin{minipage}{0.5\hsize}
     \begin{center}
    \includegraphics[width=1\textwidth]{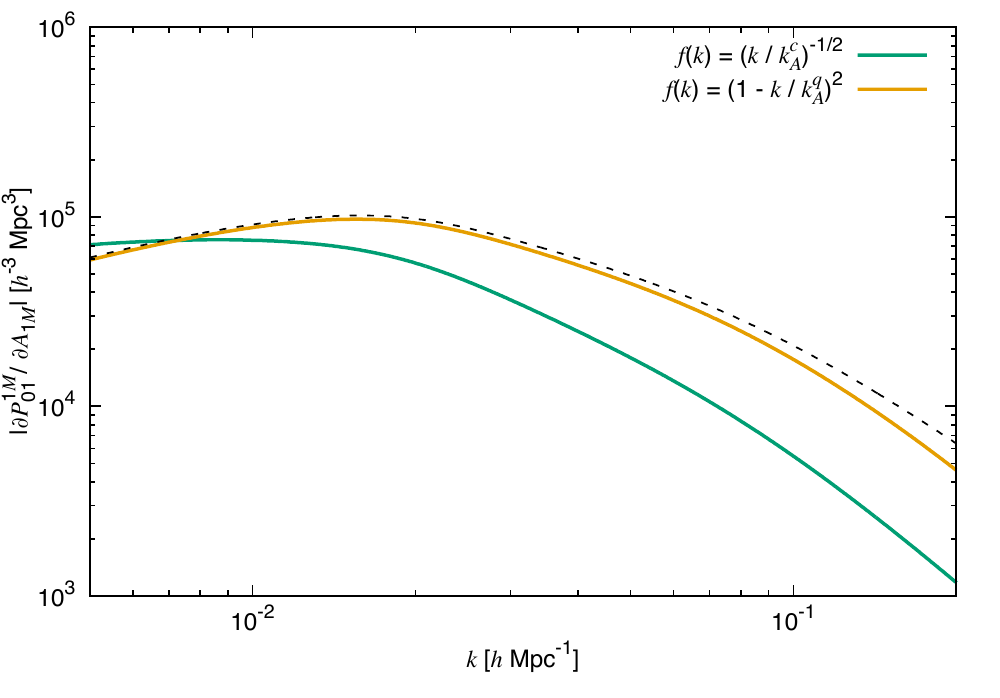}
  \end{center}
   \end{minipage}
 \end{tabular}
 \caption{Top two panels: $\partial P_{l_1 l_2}^{2M} / \partial g_{2M}$ with $f(k) = 1$ and $\partial P_{l_1 l_2}^{1M} / \partial A_{1M}$ with $f(k) = (k / k_A^c)^{-1/2}$ for all possible $(l_1, l_2)$. Bottom two panels: $\partial P_{2 0}^{2M} / \partial g_{2M}$ and $\partial P_{0 1}^{1M} / \partial A_{1M}$ for each $f(k)$. The dashed lines in the top (bottom) two panels describe $P_0$, $P_2$ and $P_4$ ($P_0$). These correspond to the results for $b = 2.0$ and $z = 0.5$. The cross point of the colored lines in the left (right) bottom panel corresponds to the pivot scale $k_g$ ($k_A^{c}$).}
\label{fig:Pl1l2}
\end{figure*}

%+++++++++++++++++++++++++++++++++++++++++++++++++++++++++++++++++++++++++
\subsection{Signatures of primordial quadrupolar asymmetry}
%+++++++++++++++++++++++++++++++++++++++++++++++++++++++++++++++++++++++++

The primordial quadrupolar power asymmetry means the primordial curvature power spectrum modulated by a directional-dependent quadratic term $g_* (\hat{k} \cdot {\hat p})^2$, where $\hat{p}$ is a preferred direction. Nonzero $g_*$ arises from anisotropic sources, such as the vector field \cite{Dimastrogiovanni:2010sm,Soda:2012zm,Maleknejad:2012fw,Bartolo:2012sd,Naruko:2014bxa,Bartolo:2014hwa,Bartolo:2015dga}, or an inflating solid or elastic medium \cite{Bartolo:2013msa, Bartolo:2014xfa}. This was tested with the observational data through a parametrization:
\begin{eqnarray}
  \Braket{\zeta_{{\bf k}_1} \zeta_{{\bf k}_2}}
  &=& (2\pi)^3 P_\zeta(k_1)  \delta^{(3)}({\bf k}_1 + {\bf k}_2) \nonumber \\ 
&& \left[ 1 + \sum_{M} g_{2M} f(k_1) Y_{2M}(\hat{k}_1) \right] ~, \label{eq:zeta2_g2M}
\end{eqnarray}
where $g_{2M}^* = (-1)^M g_{2-M}$ \cite{Ackerman:2007nb,Pullen:2007tu}. The shape of $f(k)$ depends strongly on the inflationary Lagrangian.%
\footnote{The linear growth rate $f$ should be distinguished from this $f(k)$.}
In an inflationary model involving a inflaton-vector coupling such as $G(\phi)F^2$ or $G(\phi) F \tilde{F}$, the scaling of $f(k)$ is determined by the shape of $G(\phi)$ \cite{Bartolo:2012sd,Bartolo:2014hwa,Bartolo:2015dga}. In this paper, we assume a power-law shape, $f(k) = (k/k_g)^n$ with $k_g \equiv 0.05 \, \text{Mpc}^{-1}$, and analyze in terms of 4 indices: $n = -2, -1, 0$ and $1$. These were already constrained from the latest CMB data in the {\it Planck} experiment, as $|g_{2M}| \lesssim 0.01$ \cite{Kim:2013gka,Ade:2015lrj,Ade:2015hxq}.%
\footnote{The {\it Planck} collaboration also constrained the case of $f(k) \propto k^2$, while we here do not consider this since the induced galaxy angular correlation diverges in the ultraviolet limit, spoiling the 2D Fisher matrix forecast discussed in the next section and Appendix~\ref{appen:Cl}.}
In Ref.~\cite{Pullen:2010zy}, employing the 2D galaxy angular correlation, a comparable upper bound was obtained from the photometric luminous red galaxies (LRGs) measured in the SDSS survey. 

In this case, the resultant matter power spectrum depends on $\hat{k}$, reading
\begin{eqnarray}
  P_m({\bf k}) = M_k^2 P_\zeta(k) %% \nonumber \\ 
%% &&
  \left[ 1 + \sum_{M} g_{2M} f(k) Y_{2M}(\hat{k}) \right]. \label{eq:deltam2_g2M}
\end{eqnarray}
Computing Eq.~\eqref{eq:PlldLM_def} with this by means of the law of addition of angular momentum described in Appendix~\ref{appen:math}, we obtain
 \begin{eqnarray} 
\pi_{\ell \ell'}^{LM}(k) &=& 
     P_{\ell'}(k) 
     \left[ \frac{4\pi }{2\ell+1} H_{\ell \ell 0}^{-1} \delta_{\ell, \ell'} \delta_{L,0}\delta_{M,0} \right. \nonumber \\
&&\left. + 
 \sqrt{\frac{4\pi (2\ell+1) }{(2\ell'+1) } }  H_{\ell \ell' 2}  g_{2 M} f(k) 
\delta_{L,2}
\right]. \label{eq:PlldLM_g2M} 
 \end{eqnarray}
This clearly shows that nonvanishing $\pi_{\ell \ell'}^{2M}$ can be a compelling evidence of the existence of the quadrupolar asymmetry or $g_{2M}$. A parity-even condition and a triangular inequality of $H_{\ell \ell' 2}$ restrict the allowed multipoles in $\pi_{\ell \ell'}^{2M}$ to $(\ell, \ell') = (2,0)$, $(0,2)$, $(2,2)$, $(4,2)$, $(2,4)$, $(4,4)$ and $(6,4)$.

The left top panel of Fig.~\ref{fig:Pl1l2} depicts all possible $\partial P_{\ell \ell'}^{2M} / \partial g_{2M}$ for $f(k) = 1$, where $P_{\ell \ell'}^{2M}$ is given by Eq.~\eqref{eq:pi2P}. We confirm there a magnitude relation $|P_{20}^{2M}| > |P_{02}^{2M}| \sim |P_{22}^{2M}| \sim |P_{42}^{2M}| > |P_{2 4}^{2M}| \sim |P_{4 4}^{2M}| \sim |P_{6 4}^{2M}|$, indicating that $P_{20}^{2M}$ contributes dominantly to the signal-to-noise ratio. However, $\partial P_{20}^{2M} / \partial g_{2M}$ always falls below the size of cosmic variance, $P_0$. The left bottom panel shows $P_{20}^{2M}$ for each $f(k)$. It is apparent that the tilt of $P_{20}^{2M}$ varies corresponding to the tilt of $f(k)$. Owing to this, $\partial P_{20}^{2M} / \partial g_{2M}$ for $f(k) = (k/k_g)^{-2, -1}$ and $f(k) = k/k_g$ can exceed the level of cosmic variance at $k \lesssim k_g$ and $k \gtrsim k_g$, respectively.

%+++++++++++++++++++++++++++++++++++++++++++++++++++++++++++++++++++++++++
\subsection{Signatures of primordial dipolar asymmetry}
%+++++++++++++++++++++++++++++++++++++++++++++++++++++++++++++++++++++++++

In a statistically-isotropic CMB field $T(\hat{x})$, a $3\sigma$-level directional-dependent dipolar modulation, ${\bf A} \cdot \hat{x}$, was discovered \cite{Ade:2013nlj,Akrami:2014eta,Ade:2015lrj,Ade:2015hxq,Aiola:2015rqa}. The constraints from high-$\ell$ data indicate a decaying behavior as $|{\bf A}| \simeq 0.03 (\ell/60)^{-1/2}$ \cite{Aiola:2015rqa}. There is another constraint from the density of quasars at lower redshifts, indicating the disappearance of ${\bf A}$ at $k \sim 1 \, \text{Mpc}^{-1}$ \cite{Hirata:2009ar}. This can be interpreted as the consequence of the position-dependent dipolar asymmetry generated in the primordial curvature perturbation, which is expressed as
\begin{eqnarray}
  \zeta_{\bf k}({\bf x}) = \bar{\zeta}_{\bf k}
\left[1 + \sum_M A_{1M} f(k) Y_{1M}(\hat{x}) \right] ~, \label{eq:zeta_A1M}
\end{eqnarray}
where $A_{1M}^* = (-1)^M A_{1 -M}$, and $\bar{\zeta}$ is the isotropic and homogeneous component of the curvature perturbation, whose power spectrum is given by $\Braket{\bar{\zeta}_{{\bf k}_1} \bar{\zeta}_{{\bf k}_2}} = (2\pi)^3 P_\zeta(k_1) \delta^{(3)}({\bf k}_1 + {\bf k}_2)$. A decaying behavior observed in CMB is realized by choosing $f(k) = (k / k_A^c)^{-1/2}$ with $k_A^c \equiv 0.005 \, \text{Mpc}^{-1}$. Such a decaying dipolar asymmetry can be produced e.g. by supercurvature modes of a scalar spectator field \cite{Erickcek:2008sm,Erickcek:2009at,Kanno:2013ohv,Byrnes:2016uqw}. On the other hand, there is a theoretical model with a modulation of the scalar spectral index, generating the dipolar asymmetry that vanishes at $k \sim k_A^{q} \equiv 1 \, \text{Mpc}^{-1}$ but presents at larger and smaller scales \cite{Dai:2013kfa}. In Ref.~\cite{Shiraishi:2016omb}, the detectability of the induced 21-cm power spectrum was studied by use of a parametrization $f(k) = (1-k/k_A^q)^2$. Here, let us analyze on these two theoretically- and observationally-motivated shapes.%
\footnote{
We here do not discuss $f(k) = 1 - k/k_A^q$, which was also considered in Ref.~\cite{Shiraishi:2016omb}, because the result is very similar to that for $f(k) = (1-k/k_A^q)^2$. 
}
\footnote{Strictly speaking, the curvature power spectra generated in such models break statistical homogeneity, inducing nonvanishing off-diagonal modes ${\bf k}_1 \neq -{\bf k}_2$. However, such modes are subdominant compared to the diagonal ones (because of the smallness of the departure from homogeneity) and thus the curvature perturbation (approximately) follows Eq.~\eqref{eq:zeta_A1M}.}

Assuming $|A_{1M}f(k)| \ll 1$ at observed scales, the induced matter power spectrum (under the local plane parallel approximation) depends on $\hat{x}_1 = \hat{x}_2 \equiv \hat{n}$ according to
\begin{equation}
  P_m({\bf k}) \simeq M_k^2 P_\zeta(k) 
\left[1 + 2 \sum_M A_{1M} f(k) Y_{1M}(\hat{n})  \right].
\end{equation}
The galaxy power spectrum estimated by plugging this into Eq.~\eqref{eq:deltag2} is translated into $\pi_{\ell \ell'}^{LM}$ according to Eq.~\eqref{eq:PlldLM_def}. Adding angular momenta in the Wigner symbols as done in Appendix~\ref{appen:math}, we obtain
\begin{eqnarray}
\pi_{\ell\ell'}^{LM}(k) &=& P_{\ell}(k)
     \left[  \frac{4\pi }{ 2\ell+1 } H_{\ell \ell 0}^{-1}
    \delta_{\ell, \ell'} \delta_{L,0} \delta_{M,0} \right. \nonumber \\ 
    && \left. - 2 \sqrt{\frac{4\pi (2\ell'+1) }{ (2\ell+1) } }  H_{\ell \ell' 1} A_{1 M} f(k)     
   \delta_{L,1} 
   \right]. \label{eq:PlldLM_A1M}
\end{eqnarray}
It is obvious here that the modulation due to the dipolar asymmetry creates nonvanishing $\pi_{\ell \ell'}^{1M}$ for $(\ell, \ell') = (0,1)$, $(2,1)$, $(2,3)$, $(4,3)$ and $(4,5)$. 

The right top panel of Fig.~\ref{fig:Pl1l2} displays the magnitude relation of all nonvanishing components of $\partial P_{\ell \ell'}^{1M} / \partial A_{1M}$ for $f(k) = (k/k_A^c)^{-1/2}$, where $P_{\ell \ell'}^{1M}$ is given by Eq.~\eqref{eq:pi2P}, indicating the dominance of $P_{0 1}^{1M}$. The difference of the shapes between $P_{0 1}^{1M}$ for $f(k) = (k/k_A^c)^{-1/2}$ and that for $f(k) = (1 - k/k_A^q)^2$ is seen in the right bottom panel. The inferiority of $\partial P_{0 1}^{1M} / \partial A_{1M}$ for $f(k) = (k/k_A^c)^{-1/2}$ at $ k \gtrsim k_A^c$ with respect to the level of cosmic variance $P_0$ is confirmed.

%%%%%%%%%%%%%%%%%%%%%%%%%%%%%%%%%%%%%%%%%%%%%%%%%%%%%%%%%%%%%%%%%%%%%%%%%%%%%
\section{Fisher forecasts}\label{sec:Fish}
%%%%%%%%%%%%%%%%%%%%%%%%%%%%%%%%%%%%%%%%%%%%%%%%%%%%%%%%%%%%%%%%%%%%%%%%%%%%%

\begin{figure*}[t!]
  \begin{tabular}{cc}
    \begin{minipage}{0.5\hsize}
      \begin{center}
        \includegraphics[width=1\textwidth]{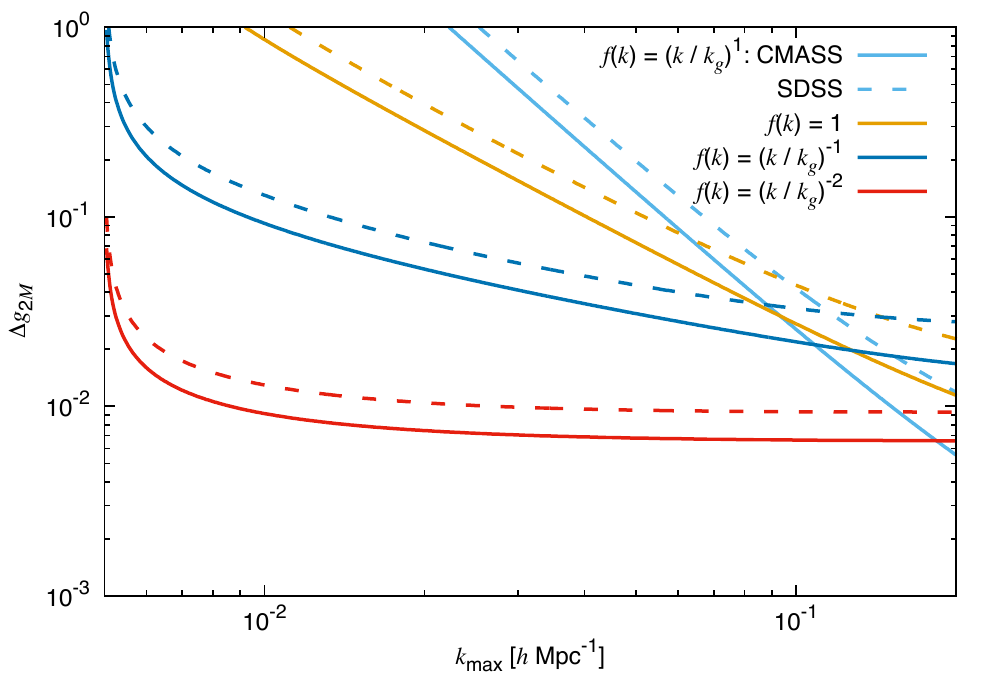}
  \end{center}
    \end{minipage}
    \begin{minipage}{0.5\hsize}
      \begin{center}
         \includegraphics[width=1\textwidth]{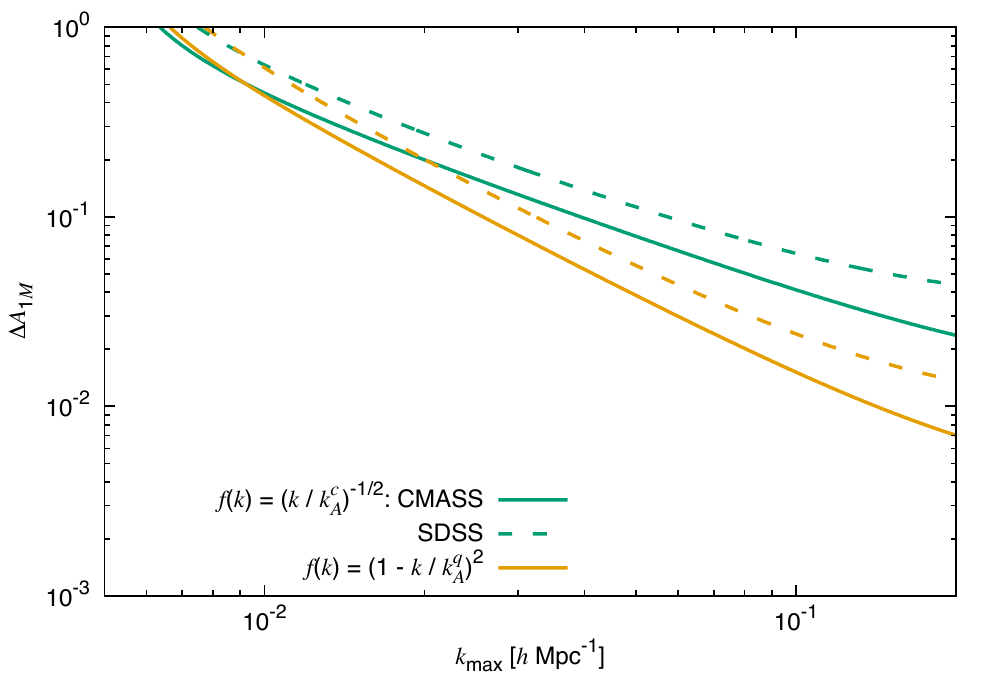}
  \end{center}
\end{minipage}
    \end{tabular}
\\
 \begin{tabular}{cc}
    \begin{minipage}{0.5\hsize}
      \begin{center}
         \includegraphics[width=1\textwidth]{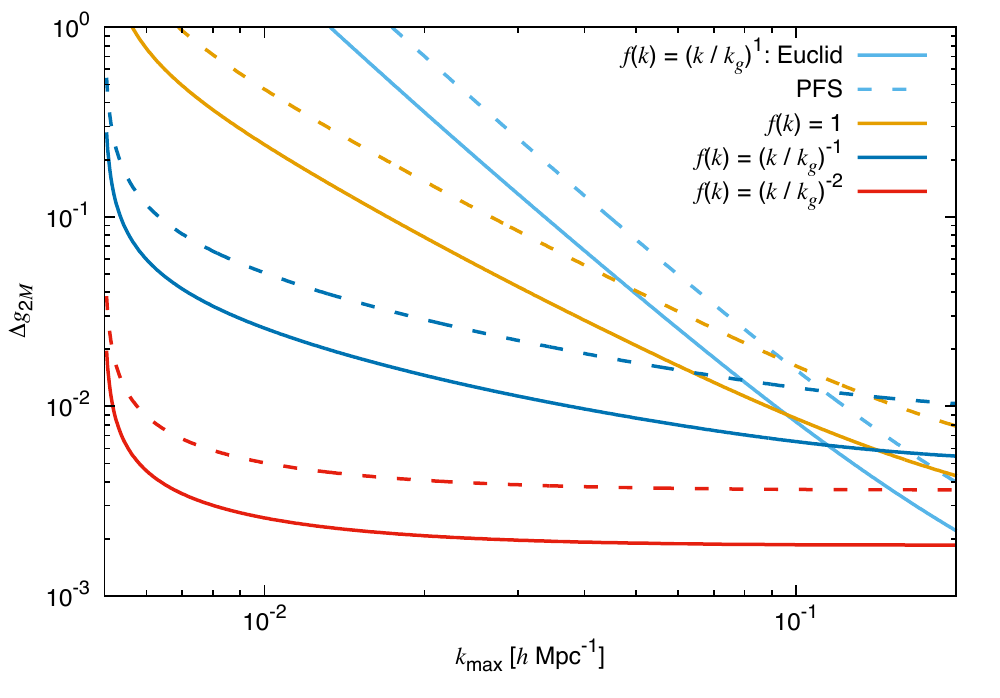}
  \end{center}
    \end{minipage}
    \begin{minipage}{0.5\hsize}
      \begin{center}
         \includegraphics[width=1\textwidth]{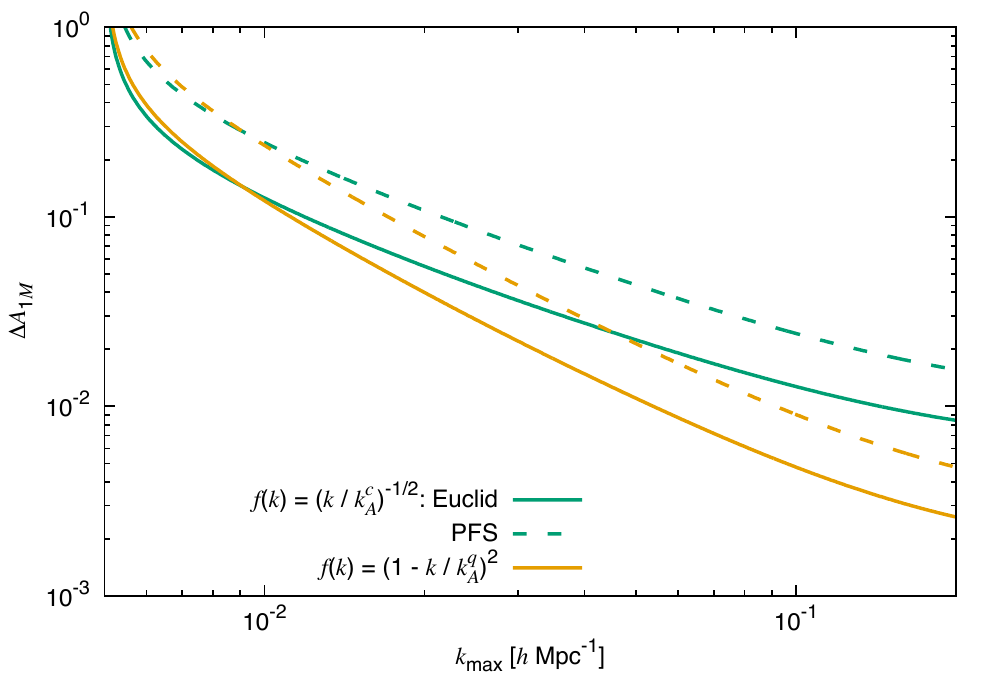}
  \end{center}
\end{minipage}
 \end{tabular}
 \caption{Expected $1\sigma$ errors $\Delta g_{2M}$ (left panels) and $\Delta A_{1M}$ (right panels) for each $f(k)$, as a function of $k_{\rm max}$. These are obtained from the 3D Fisher matrix \eqref{eq:Fish3D} assuming the SDSS-like, CMASS-like (top panels), PFS-like and Euclid-like (bottom panels) surveys. We adopt $k_{\rm min} = 0.005 h \, \text{Mpc}^{-1}$.}\label{fig:error_g2M_A1M_3D_vs_2D} 
\end{figure*}

In this section, we estimate the sensitivities to $g_{2M}$ and $A_{1M}$. The Fisher matrix for $h_{LM} \ni g_{2M}, A_{1M}$ from the reduced BipoSH coefficients $P_{\l_1 l_2}^{LM} (\equiv {\bf P})$ \eqref{eq:pi2P} is defined by
\begin{eqnarray}
   {\bf F} &=& \frac{\partial {\bf P}}{\partial {\bf h}} 
   {\rm \bf Cov}^{-1}[{\bf P}^\dagger, {\bf P}]
  \left(\frac{\partial {\bf P}}{\partial {\bf h}} \right)^\dagger ~ ,
\end{eqnarray}
where we have adopted a matrix representation. Via Eq.~\eqref{eq:PlldLM_def}, the covariance ${\rm \bf Cov}[{\bf P}^\dagger, {\bf P}] = \Braket{P_{l_1 l_2}^{LM *}(k) P_{l_1' l_2'}^{L' M'}(k') }_c$ is translated from that of $P^s$,
\begin{widetext}
\begin{eqnarray}
\Braket{P^s({\bf k},\hat{n}) P^s({\bf k}',\hat{n}')}_c 
  = 4\pi \frac{\delta_{k,k'}}{N_k}  \left[ \sum_{J} P_J^{(\rm O)}(k) {\cal L}_J(\hat{k} \cdot \hat{n}) \right]^2 %% \nonumber \\
%% &&\quad \times
  \left[ \delta^{(2)}(\hat{k} + \hat{k}') + \delta^{(2)}(\hat{k} - \hat{k}') \right] %% \nonumber \\
%% &&
  4\pi \delta^{(2)}(\hat{n} - \hat{n}')~,
\end{eqnarray}
where $\Braket{\cdots}_c$ is the connected part of the ensemble average, and $N_k = 4\pi k^2 \Delta k V / (2\pi)^3$ is the mode number with $V$ and $\Delta k$ being the volume of the survey area and the interval between each Fourier mode. The Legendre coefficients are given by the sum of cosmic variance and the homogeneous shot noise and hence $ P_0^{(\rm O)} = P_0 + 1 / n_g$, $P_{2}^{(\rm O)} = P_{2}$, $P_{4}^{(\rm O)} = P_{4}$ and $P_{1}^{(\rm O)} = P_{3}^{(\rm O)} = P_{J \geq 5}^{(\rm O)} = 0$ with $n_g$ denoting the number density of galaxies. Since our analysis is based on the local plane parallel approximation, the covariance may include $4\pi \delta^{(2)}(\hat{n} - \hat{n}')$. Adding angular momenta by following the computational procedure in Appendix~\ref{appen:math}, we can derive
\begin{eqnarray}
\Braket{P_{l_1 l_2}^{LM *}(k) P_{l_1' l_2'}^{L' M'}(k') }_c
  = \delta_{L, L'}\delta_{M, M'}   \frac{\delta_{k,k'}}{N_k} \Theta_{l_1 l_2, l_1' l_2'}^L(k) ~,
\end{eqnarray}
where
\begin{eqnarray}
  \Theta_{l_1 l_2, l_1' l_2'}^L(k) 
  &=& (2l_1 + 1)(2l_2 + 1)(2l_1' + 1)(2l_2' + 1)%%  \nonumber \\
 %% && 
 (2L+1)  (-1)^{l_1} \left[1 + (-1)^{l_1'}\right]
  H_{l_1 l_2 L} H_{l_1' l_2' L} \sum_{J J'} P_J^{(\rm O)}(k) P_{J'}^{(\rm O)}(k)
  \nonumber \\
  && 
   \sum_{L_1 L_2}  
   (2L_1 + 1) (2L_2 + 1) %% \nonumber \\ 
  %% &&
  H_{l_1 J L_1} H_{l_2 J L_2} H_{l_1' J' L_1} H_{l_2' J' L_2}
  \left\{
    \begin{smallmatrix}%{array}{ccc}
      L & L_1 & L_2\\
      J & l_2 & l_1 
    \end{smallmatrix}
    \right\}
    \left\{
    \begin{smallmatrix}%{array}{ccc}
      L & L_1 & L_2\\
      J' & l_2' & l_1' 
    \end{smallmatrix}%{array}
    \right\} ~,
\end{eqnarray}
with the function enclosed by the curly bracket denoting the Wigner $6j$ symbol.
The Fisher matrix is therefore diagonalized as
\begin{eqnarray}
F_{LM, L' M'} &=& \delta_{L, L'} \delta_{M, M'} V \int_{k_{\rm min}}^{k_{\rm max}} \frac{k^2 dk}{ 2\pi^2} \sum_{l_1 l_2 l_1' l_2'} %%  \nonumber \\
  %% &&
 \frac{\partial P_{l_1 l_2}^{L M}(k) }{\partial h_{L M}} \left(\Theta^{-1}\right)_{l_1 l_2, l_1' l_2'}^{L}(k)
  \left( \frac{\partial P_{l_1' l_2'}^{L M}(k) }{\partial h_{L M}} \right)^* ~, \label{eq:Fish3D}
\end{eqnarray}
where we have taken the continuous limit $\sum_k \Delta k \to \int dk$. This is the expression for one redshift bin, while the co-add information from $N_{\rm bin}$ independent redshift bins further enhances the Fisher matrix as
\end{widetext}
\begin{eqnarray}
  F_{LM, LM}^{\rm tot} = \sum_{i=1}^{N_{\rm bin}} F_{LM, LM}(z_i) ~. \label{eq:Fish_tomography}
\end{eqnarray}
The expected $1\sigma$ errors on $ h_{LM}$ are computed according to $\Delta h_{LM} = 1 / \sqrt{F_{LM,LM}^{\rm tot}}$.%
\footnote{
We here ignore the information from the cross-correlated power spectra between different redshifts for simplicity, while adding them, in principle, further improves the sensitivity.
}

\begin{table}[t!]
      \begin{tabular}{|c||c|c|c|c|}
  \hline
  &  $z$ & $b$ & $n_g$ [$h^3 \, \text{Mpc}^{-3}$] & $V$ [$h^{-3} \, \text{Gpc}^{3}$] \\ \hline \hline
  SDSS \cite{Abazajian:2008wr} & 0.33 & 2.00 & $6.7 \times 10^{-5}$ & 1.58 \\ \hline
  %------
  CMASS \cite{Eisenstein:2011sa} & 0.50 & 2.00 & $2.9 \times 10^{-4}$ & 2.50 \\ \hline
  %------
  PFS \cite{Ellis:2012rn} & 0.70 & 1.18 & $1.9 \times 10^{-4}$ & 0.59 \\ % \hline
  & 0.90 & 1.26 & $6.0 \times 10^{-4}$ & 0.79 \\ % \hline
  & 1.10 & 1.34 & $5.8 \times 10^{-4}$ & 0.96 \\ % \hline
  & 1.30 & 1.42 & $7.8 \times 10^{-4}$ & 1.09 \\ % \hline
  & 1.50 & 1.50 & $5.5 \times 10^{-4}$ & 1.19 \\ % \hline
  & 1.80 & 1.62 & $3.1 \times 10^{-4}$ & 2.58 \\ % \hline
  & 2.20 & 1.78 & $2.7 \times 10^{-4}$ & 2.71 \\ \hline
  %------
  Euclid  & 1.15 & 1.33 & $6.1 \times 10^{-4}$ & 4.57 \\ % \hline
  \cite{Laureijs:2011gra,Spergel:2013tha} & 1.25 & 1.38 & $5.2 \times 10^{-4}$ & 4.84 \\ % \hline
  & 1.35 & 1.44 & $4.4 \times 10^{-4}$ & 5.08 \\ % \hline
  & 1.45 & 1.50 & $3.5 \times 10^{-4}$ & 5.28 \\ % \hline
  & 1.55 & 1.55 & $2.5 \times 10^{-4}$ & 5.45 \\ % \hline
  & 1.65 & 1.61 & $1.6 \times 10^{-4}$ & 5.59 \\ % \hline
  & 1.75 & 1.67 & $1.5 \times 10^{-4}$ & 5.71 \\
  & 1.85 & 1.73 & $9.3 \times 10^{-5}$ & 5.80 \\ \hline
  \end{tabular} 
      \caption{Experimental parameters adopted for computations of the Fisher matrix~\eqref{eq:Fish3D}.}
  \label{tab:survey}
\end{table}

Figure~\ref{fig:error_g2M_A1M_3D_vs_2D} describes $\Delta g_{2M}$ and $\Delta A_{1M}$ obtained from the above Fisher matrix as a function of $k_{\rm max}$. We are now interested in how much the sensitivity to $g_{2M}$ or $A_{1M}$ can improve as time goes by and thus analyze with $V$, $n_g$ and $b$ in four generation surveys: SDSS \cite{Abazajian:2008wr}, BOSS CMASS \cite{Eisenstein:2011sa}, PFS \cite{Ellis:2012rn} and Euclid \cite{Laureijs:2011gra,Spergel:2013tha}. The numbers adopted here are listed in Table~\ref{tab:survey}. Assuming that the cosmological parameters are well-determined by analyzing the isotropic component $P_{\ell \ell}^{00}$, we fix them to be the values close to the current observational limits \cite{Ade:2015xua} and do not vary in this Fisher matrix computation. Compared with SDSS or CMASS, $V$ and $n_g$ will drastically increase in the future surveys as PFS and Euclid. This fact could result in the sensitivity beyond the {\it Planck} results ($\Delta g_{2M} \sim  \Delta A_{1M} \sim 10^{-2}$ \cite{Kim:2013gka,Ade:2015lrj,Ade:2015hxq,Aiola:2015rqa}) at high $k_{\rm max}$, as shown in the bottom panels.

To understand the $k_{\rm max}$ dependence, let us find analytic expressions of the Fisher matrix. Owing to the largeness of $P_{20}^{2M}$ or $P_{01}^{1M}$ compared with the others (as seen in the previous section), we may evaluate the Fisher matrix with such a dominant component alone. This treatment is reasonable because the contribution from the others is less than sub percent of that from $P_{20}^{2M}$ or $P_{01}^{1M}$. Using a fact that $P_0^{(\text{O})} \gg P_{J \geq 1}^{(\text{O})}$, we find the reduced expressions:
\begin{eqnarray}
F_{2M, 2M}^{(g)} &\simeq&  V \int_{k_{\rm min}}^{k_{\rm max}} \frac{k^2 dk}{ 2\pi^2} \frac{f^2 (k)}{8 \pi } 
  \left( \frac{P_{0}(k) }{P_0^{(\rm O)}(k)} \right)^2 ~, \\
  %--------
F_{1M, 1M}^{(A)}  &\simeq& V \int_{k_{\rm min}}^{k_{\rm max}} \frac{k^2 dk}{ 2\pi^2}
 \frac{f^2(k)}{2\pi }
 \left( \frac{P_{0}(k)}{P_0^{(\rm O)}(k)} \right)^2 ~.
\end{eqnarray}
Let us consider a cosmic-variance-limited (CVL) level measurement. The shot noise is then ignored, i.e. $P_{0}(k) / P_0^{(\rm O)}(k) \simeq 1$, and thus the Fisher matrix is independent of $ n_g$. By assuming $f(k) = (k / k_g)^n$ or $f(k) = (k / k_A)^n$, the $k$ integrals are computed analytically and hence we obtain
%\begin{widetext}
\begin{eqnarray}
&&  F_{2M, 2M}^{(g) \rm tot} \simeq \frac{V_{\rm tot} k_g^3}{16\pi^3 } \nonumber \\ 
&&  \times
  \begin{cases}
      \left[(\frac{k_{\rm max}}{k_g})^{2n+3} - (\frac{k_{\rm min}}{k_g})^{2n+3} \right] \frac{1}{2n+3}   &: n \neq -\frac{3}{2} \\
      \ln(\frac{k_{\rm max}}{k_{\rm min}})   &: n = -\frac{3}{2}
    \end{cases}, \label{eq:Fish3D_g2M_CV} \\ 
    %-----
&& F_{1M,1M}^{(A) \rm tot} \simeq \frac{V_{\rm tot} k_A^3}{4\pi^3 } \nonumber \\ 
&& \times  
  \begin{cases}
    \left[(\frac{k_{\rm max}}{k_A})^{2n+3} - (\frac{k_{\rm min}}{k_A})^{2n+3} \right] \frac{1}{2n+3}   &: n \neq -\frac{3}{2} \\
    \ln(\frac{k_{\rm max}}{ k_{\rm min}})   &: n = -\frac{3}{2}
  \end{cases}. \label{eq:Fish3D_A1M_CV}
\end{eqnarray}
%\end{widetext}
where $V_{\rm tot} = \sum_{i=1}^{N_{\rm bin}}V(z_i)$ is the total survey volume. These indicate that the Fisher matrix is sensitive to $k_{\rm max}$ or $k_{\rm min}$ when $n > -3/2$ or $n < -3/2$. The $1\sigma$ error bars computed from these recover the lines in the SDSS case in Fig.~\ref{fig:error_g2M_A1M_3D_vs_2D}, and especially agree with the lines in the CMASS, PFS or Euclid case because of the smallness of the shot noise. Note that the lines for the dipolar asymmetry case with $f(k) = (1 - k / k_A^q)^2$ are also recovered by Eq.~\eqref{eq:Fish3D_A1M_CV} with $n = 0$, since $f(k)$ is almost flat within $k_{\rm max}$ shown in Fig.~\ref{fig:error_g2M_A1M_3D_vs_2D}.

\begin{table}[t]
  \begin{tabular}{|c||c|c|c|c|}
  \hline
  $f(k)$ & SDSS & CMASS & PFS & Euclid \\ \hline\hline
  %--- pow 1
  $(k/k_g)^1$ & $1.2$ ($3.3$) & $0.55$ ($2.4$) & $0.40$ & $0.22$ \\
  %--- pow 0
  $1$ & $2.3$ ($5.1$) & $1.1$ ($3.5$) & $0.78$ & $0.43$ \\
  %--- pow -1
  $(k/k_g)^{-1}$ & $2.8$ ($3.5$) & $1.7$ ($2.5$) & $1.0$ & $0.55$ \\
  %--- pow -2
  $(k/k_g)^{-2}$ & $0.93$ ($0.65$) & $0.66$ ($0.51$) & $0.36$ & $0.19$ \\
  \hline
  \end{tabular}
  \caption{Expected $1\sigma$ errors $\Delta g_{2M} / 10^{-2}$ estimated from the 3D Fisher matrix at $k_{\rm max} = 0.2 h \, \text{Mpc}^{-1}$. For the SDSS (CMASS) case, we also present the results from the 2D Fisher matrix at the corresponding angular resolution, i.e. $\ell_{\rm max} \simeq 280$ ($350$), in the brackets. We here adopt $k_{\rm min} = 0.005 h \, \text{Mpc}^{-1}$ and $\ell_{\rm min} = 2$.}
  \label{tab:g2M}
\end{table}

\begin{table}[t]
\begin{tabular}{|c||c|c|c|c|}
  \hline
  $f(k)$ & SDSS & CMASS & PFS & Euclid \\ \hline\hline
  %--- pow -1/2
  $(k/k_A^c)^{-1/2}$ & $4.4$ ($7.2$) & $2.4$ ($5.0$) & $1.6$ & $0.84$ \\
  %--- dip
  $(1 - k/k_A^q)^2$ & $1.4$ ($2.9$) & $0.70$ ($2.0$) & $0.48$ & $0.26$ \\
  \hline
\end{tabular}
\caption{Expected $1\sigma$ errors $\Delta A_{1M} / 10^{-2}$, but otherwise the same parameters as Table~\ref{tab:g2M}.}
\label{tab:A1M}
\end{table}

Tables~\ref{tab:g2M} and \ref{tab:A1M} summarize $\Delta g_{2M}$ and $\Delta A_{1M}$ at $k_{\rm max} = 0.2 h \, \text{Mpc}^{-1}$, respectively, for each survey. For comparison, in the SDSS and CMASS cases, we present the results obtained from the 2D Fisher matrix \eqref{eq:Fish2D_g2M} or \eqref{eq:Fish2D_A1M} at the corresponding angular resolutions, which are given by $\ell_{\rm max} = k_{\rm max} x(\bar{z})$ with $\bar{z} = 0.4$ and $0.5$ being mean redshifts in SDSS and CMASS, respectively.%
\footnote{The experimental information required for the 2D Fisher matrix analysis (i.e. the shape of the radial selection function $W(z)$, the number of galaxies distributed on a 2D observed region $N_g$ and the fraction of the sky coverage $f_{\rm sky}$) is adopted from Refs.~\cite{Padmanabhan:2006cia,Pullen:2010zy} (SDSS) and \cite{Ross:2011cz,Ho:2012vy} (CMASS).} 
This table indicates that, except for $f(k) \propto k^{-2}$, the 3D estimator could outperform the 2D one. This is due to a fact that, if $f(k)$ is blue-tilted compared with $k^{-3/2}$, $\Delta h_{LM}^{\rm 3D}$ decreases more rapidly than $\Delta h_{LM}^{\rm 2D}$ as $k_{\rm max}$ or $ \ell_{\rm max}$ increases, e.g., $\Delta h_{LM}^{\rm 3D} \propto k_{\rm max}^{-3/2}$  vs. $\Delta h_{LM}^{\rm 2D} \propto \ell_{\rm max}^{-1}$ for the $f(k)= 1$ case [see Eqs.~\eqref{eq:Fish3D_g2M_CV}, \eqref{eq:Fish3D_A1M_CV}, \eqref{eq:Fish2D_g2M_CV} and \eqref{eq:Fish2D_A1M_CV}].

%%%%%%%%%%%%%%%%%%%%%%%%%%%%%%%%%%%%%%%%%%%%%%%%%%%%%%%%%%%%%%%%%%%%%%%%%%%%%
\section{Conclusions}\label{sec:conclusion}
%%%%%%%%%%%%%%%%%%%%%%%%%%%%%%%%%%%%%%%%%%%%%%%%%%%%%%%%%%%%%%%%%%%%%%%%%%%%%

Statistical isotropy is an essential property of the cosmological fluctuations; hence, their accurate tests are required for unambiguous establishment of the cosmological scenario. In this paper, we developed an efficient way to test isotropy with observed galaxy distributions. An estimator based on the 2D angular correlation function had been discussed in the earlier literature, while we here considered that based on the 3D correlation function for the first time.

If the Universe is statistically anisotropic, in addition to $\hat{k} \cdot \hat{n}$ due to redshift-space distortions, extra angular dependence arises in the galaxy power spectrum measured from galaxy surveys. The usual Legendre expansion is then no longer valid; thus, we performed a new type of decomposition using the BipoSH basis \eqref{eq:S_basis}. In the resultant expansion coefficients $\pi_{\ell \ell'}^{LM}$, the $L \neq 0$ modes can never generated unless the isotropic condition is violated, so they are unbiased observables of the symmetry breakings.

For a demonstration, we considered a situation that the power spectrum of primordial curvature perturbations contains the usual quadrupolar-type ($g_{2M}$) or dipolar-type ($A_{1M}$) power asymmetry, and computed $\pi_{\ell \ell'}^{LM}$ from the induced galaxy power spectrum. We then confirmed that the quadrupolar and dipolar modulations create nonvanishing $L = 2$ and $L=1$ components, respectively.

The signal-to-noise ratio of the 3D correlation function is more sensitive to the number of available Fourier modes than the 2D case. Owing to this, our 3D estimator outperforms the usual 2D one in many cases. The Fisher matrix computations based on $\pi_{\ell \ell'}^{LM}$ led to a ${\it Planck}$-level sensitivity $\Delta g_{2M} \sim \Delta A_{1M} \sim 0.01$ in the measurement with the SDSS or CMASS data. The error bars could shrink by an order of magnitude in the futuristic surveys as PFS and Euclid.

Besides the primordial symmetry breakings, the statistically anisotropic signal could also arise from some late-time phenomena \cite{Pontzen:2010eg} or the super sample signal beyond the survey area \cite{Takada:2013bfn,Li:2014jra,Ip:2016jji,Akitsu:2016leq}. These other sources could also generate some other nontrivial features in the $L \neq 0$ components of $\pi_{\ell \ell'}^{LM}$, motivating further theoretical investigations. The application of our decomposition technique to the real data is another interesting and important topic. Toward this end, nontrivial effects from small-scale nonlinear physics \cite{Ando:2008zza} and contaminations due to specific survey geometry need to be studied. These theoretical and observational issues will be addressed in our ongoing and forthcoming projects.

%%%%%%%%%%%%%%%%%%%%%%%%%%%%%%%%%%%%%%%%%%%%%%%%%%%%%%%%%%%%%%%%%%%%%%%%%%%%%

\acknowledgements

We thank Masahiro Takada for fruitful discussions. M.\,S. and N.\,S.\,S. were supported in part by a Grant-in-Aid for JSPS Research under Grant Nos.~27-10917 and 28-1890, respectively. T.\,O. was supported by JSPS KAKENHI Grant Number JP26887012. We were supported in part by the World Premier International Research Center Initiative (WPI Initiative), MEXT, Japan. Numerical computations were in part carried out on Cray XC30 at Center for Computational Astrophysics, National Astronomical Observatory of Japan.

%%%%%%%%%%%%%%%%%%%%%%%%%%%%%%%%%%%%%%%%%%%%%%%%%%%%%%%%%%%%%%%%%%%%%%%%%%%%%%%%%%%%%%%%%%%%%%%%%%%%%%%%%%%%%%%%%%%%%%%%%%%%%%%%%%%%%%%%%%%%%%%%%%%%%%%%%%%%%%%%%%%%%%%%%%%%%%
\appendix

%%%%%%%%%%%%%%%%%%%%%%%%%%%%%%%%%%%%%%%%%%%%%%%%%%%%%%%%%%%%%%%%%%%%%%%%%%%%%
\section{Angular correlations from the primordial quadrupolar and dipolar asymmetries}\label{appen:Cl}
%%%%%%%%%%%%%%%%%%%%%%%%%%%%%%%%%%%%%%%%%%%%%%%%%%%%%%%%%%%%%%%%%%%%%%%%%%%%%

In this Appendix we formulate the 2D angular power spectrum and the Fisher matrix that are used for the comparison with the 3D results in Sec.~\ref{sec:Fish}. A similar formalism was already discussed in the literature of the galaxy \cite{Pullen:2010zy}, CMB \cite{Ackerman:2007nb,Pullen:2007tu,Hanson:2009gu} and 21-cm \cite{Shiraishi:2016omb} power spectra, so we do not present the detailed derivations.

Now we move to 2D harmonic space according to 
\begin{eqnarray}
  a_{\ell m} = \int d^2 \hat{x} Y_{\ell m}^*(\hat{x}) \int_0^{\infty} dz W(z) \delta^s({\bf x})  ~, \label{eq:alm_def} 
\end{eqnarray}
where $W(z)$ is a radial selection function satisfying $\int_0^{\infty} dz W(z) = 1$, extracting the information of galaxy distributions at each redshift slice. The shape strongly depends on the design of the galaxy survey under examination. In the $g_{2M}$-type asymmetry case, the harmonic coefficient is formulated as
\begin{eqnarray}
  a_{\ell m} &=&
  4\pi i^\ell 
  \int \frac{d^3 k}{(2\pi)^3} Y_{\ell m}^*(\hat{k}) \zeta_{\bf k} 
 {\cal T}_{\ell}(k) ~, \label{eq:alm21_iso}
  \end{eqnarray}
where the transfer function ${\cal T}_{\ell}(k)$ reads \cite{Padmanabhan:2006cia}
\begin{eqnarray}
  {\cal T}_{\ell}(k)
  &=& t_\ell(k)
  + \frac{2 \ell^2 + 2 \ell - 1}{(2 \ell - 1) (2 \ell + 3)} \tilde{t}_\ell(k) \nonumber \\
  &&
  - \frac{(\ell-1) \ell }{(2\ell - 1)(2\ell + 1)} \tilde{t}_{\ell-2}(k) \nonumber \\
  &&
  - \frac{(\ell+1) (\ell+2)}{(2 \ell+1) (2 \ell+3)} \tilde{t}_{\ell+2}(k) 
  ~,
\end{eqnarray}
with
\begin{eqnarray}
 t_\ell(k) &=& \int_0^{\infty} dz W(z)
 b(z)   M_k(z) j_\ell(k x(z)) ~, \\
 %@@@@
 \tilde{t}_\ell(k) &=& \int_0^{\infty} dz W(z) 
  f(z)  M_k(z)  j_\ell(k x(z))   ~.
  \end{eqnarray}
This and Eq.~\eqref{eq:zeta2_g2M} yield the $g_{2M}$-type angular power spectrum given by $\Braket{a_{\ell_1 m_1} a_{\ell_2 m_2}}
  =  C_{\ell_1} (-1)^{m_1} \delta_{\ell_1, \ell_2} \delta_{m_1, -m_2} %% \nonumber \\
  %% && 
  + (-1)^{m_2} C_{\ell_1 m_1, \ell_2 -m_2}$, where
\begin{eqnarray}
  C_{\ell} &=& \frac{2}{\pi}
  \int_0^\infty k^2 dk P_\zeta(k) 
{\cal T}_{\ell}^2(k) ~, \label{eq:Cl_iso} \\
  %-------
  C_{\ell_1 m_1, \ell_2 m_2}
  &=& i^{\ell_1 - \ell_2}
  G_{\ell_1 \ell_2}
  (-1)^{m_1} 
h_{\ell_1 \ell_2 2} \nonumber \\ 
&& \sum_{M}  g_{2M} 
\left(
\begin{smallmatrix}%{array}{ccc}
  \ell_1 & \ell_2 & 2 \\
   -m_1 & m_2 & M 
  \end{smallmatrix}%{array}
 \right) ~, \label{eq:Cl1l2_g2M}
\end{eqnarray}
with $h_{l_1 l_2 l_3} \equiv \sqrt{(2 l_1 + 1)(2 l_2 + 1)(2 l_3 + 1) / (4 \pi)} \left(\begin{smallmatrix}%{array}{ccc}
  l_1 & l_2 & l_3 \\
   0 & 0 & 0 
  \end{smallmatrix}\right)$ and 
\begin{eqnarray}
  G_{\ell_1 \ell_2}
 &=& \frac{2}{\pi}
  \int_0^\infty k^2 dk P_\zeta(k) f(k)
{\cal T}_{\ell_1}(k)  
{\cal T}_{\ell_2}(k) \label{eq:Gfunc} ~.
\end{eqnarray}
The existence of the geometrical function $h_{\ell_1 \ell_2 2}$ indicates that $C_{\ell_1 m_1, \ell_2 m_2}$ has nonvanishing signal at not only the diagonal part $\ell_1 = \ell_2$ but also the off-diagonal one obeying $|\ell_1 - \ell_2| = 2$.

The harmonic coefficients from the curvature perturbation involving the dipolar asymmetry \eqref{eq:zeta_A1M} can be divided into the symmetric component and the asymmetric modulation as $a_{\ell m} = \bar{a}_{\ell m} +  a_{\ell m}^{\rm dip} ~$, where $\bar{a}_{\ell m}$ is the same as Eq.~\eqref{eq:alm21_iso}, and 
\begin{eqnarray}
  a_{\ell m}^{\rm dip}
  &=&
  \sum_{LM} 4\pi i^L 
  \int \frac{d^3 k}{(2\pi)^3}
  Y_{LM}^*(\hat{k}) \bar{\zeta}_{\bf k} f(k)
  {\cal T}_L(k) \nonumber \\ 
 && 
  (-1)^m
  h_{\ell L 1}
  \sum_{M'} A_{1M'} 
  \left(
  \begin{smallmatrix}%{array}{ccc}
    \ell & L & 1 \\
    -m & M & M'
  \end{smallmatrix}%{array}
 \right).
\end{eqnarray}
Assuming $|A_{1M} f(k)| \ll 1$, we may ignore a ${\cal O}(|A_{1M} f(k)|^2)$ term and hence obtain $C_\ell$ equivalent to Eq.~\eqref{eq:Cl_iso} and 
\begin{eqnarray}
C_{\ell_1 m_1, \ell_2 m_2} 
&=&  \left( G_{\ell_1 \ell_1}
  + G_{\ell_2 \ell_2} \right) 
  (-1)^{m_1} h_{\ell_1 \ell_2 1} \nonumber \\ 
&&  \sum_{M} A_{1M} 
  \left(
  \begin{smallmatrix}%{array}{ccc}
    \ell_1 & \ell_2 & 1 \\
    -m_1 & m_2 & M
  \end{smallmatrix}%{array}
  \right) ~. \label{eq:Cl1l2_A1M}
\end{eqnarray}
One can confirm from the triangular inequality and parity-even condition of $h_{\ell_1 \ell_2 1}$ that $C_{\ell_1 m_1, \ell_2 m_2}$ can take nonzero values only when $|\ell_1 - \ell_2| = 1$ holds.

The Fisher matrix derived on the basis of the diagonal covariance matrix approximation reads \cite{Pullen:2007tu,Hanson:2009gu, Hanson:2010gu}
\begin{eqnarray}
  F_{LM, L' M'}
&=& \frac{f_{\rm sky}}{2} 
  \sum_{\ell_1 m_1 \ell_2 m_2}
  \frac{\partial C_{\ell_1 m_1 \ell_2 m_2}}{\partial h_{L M}^{*}}
     \nonumber \\ 
     && \left( C_{\ell_1}^{\rm (O)} C_{\ell_2}^{\rm (O)} \right)^{-1}
     \frac{\partial C_{\ell_1 m_1 \ell_2 m_2}^{*}}{\partial h_{L'M'}}  ~.
  \end{eqnarray}
Substituting Eqs.~\eqref{eq:Cl1l2_g2M} and \eqref{eq:Cl1l2_A1M} into this leads to  the bottom-line forms, reading
\begin{equation}
 F_{2M, 2 M'}^{(g)} = \delta_{M, M'} \frac{ f_{\rm sky}}{10} 
  \sum_{\ell_1, \ell_2 = \ell_{\rm min}}^{\ell_{\rm max}} h_{\ell_1 \ell_2 2}^2
  \frac{G_{\ell_1 \ell_2}^2}{ C_{\ell_1}^{\rm (O)} C_{\ell_2}^{\rm (O)} } ~, \label{eq:Fish2D_g2M}
\end{equation}
for the $g_{2M}$ case and 
\begin{equation}
  %----------
  F_{1M, 1 M'}^{(A)} = \delta_{M,M'}
 \frac{f_{\rm sky}}{6}
  \sum_{\ell_1, \ell_2 = \ell_{\rm min}}^{\ell_{\rm max}}
%% \nonumber \\ 
%% && 
  h_{\ell_1 \ell_2 1}^2
  \frac{\left( G_{\ell_1 \ell_1} + G_{\ell_2 \ell_2} \right)^2}{C_{\ell_1}^{\rm (O)} C_{\ell_2}^{\rm (O)}} ~, 
    \label{eq:Fish2D_A1M}
\end{equation}
for the $A_{1M}$ case. The Fisher matrix including co-add information of multi-redshift slices, $F^{\rm tot}$, is obtained via Eq.~\eqref{eq:Fish_tomography}. The expected 1$\sigma$ errors can be computed by $\Delta h_{LM} = 1/\sqrt{F_{LM, L M}^{\rm tot} }$.

If considering realistic surveys, $C_\ell^{\rm (O)}$ includes the shot noise in addition to cosmic variance, as $C_\ell^{\rm (O)} = C_\ell + 4\pi f_{\rm sky} / N_g$, where $N_g$ is the number of galaxies distributed in the survey area on 2D sphere. On the other hand, assuming a noiseless CVL survey where $C_\ell^{\rm (O)} \simeq C_\ell$ holds, the Fisher matrix with $f(k) \simeq 1$ simplifies to
\begin{eqnarray}
 F_{2M, 2 M}^{(g) f(k) \simeq 1} &\simeq& \frac{ f_{\rm sky}}{10} 
  \sum_{\ell_1, \ell_2 = \ell_{\rm min}}^{\ell_{\rm max}} h_{\ell_1 \ell_2 2}^2 ~, \\
  %----------
  F_{1M, 1 M}^{(A) f(k) \simeq 1} &\simeq& 
 \frac{2f_{\rm sky}}{3}
  \sum_{\ell_1, \ell_2 = \ell_{\rm min}}^{\ell_{\rm max}} 
  h_{\ell_1 \ell_2 1}^2
   ~, 
\end{eqnarray}
where we have used a fact that $G_{\ell_1 \ell_2} \simeq C_{\ell_1}$ within allowed ranges of $\ell_1$ and $\ell_2$. By choosing $\ell_{\rm max} \gg \ell_{\rm min}$, the summations over $\ell_1$ and $\ell_2$ are analytically evaluated \cite{Shiraishi:2016omb} and we thus derive
\begin{eqnarray}
 F_{2M, 2 M}^{(g) f(k) \simeq 1} &\simeq& \frac{f_{\rm sky}}{8\pi}  \ell_{\rm max}^{2} ~, \label{eq:Fish2D_g2M_CV} \\ 
  %----
 F_{1M, 1 M}^{(A) f(k) \simeq 1} &\simeq& \frac{f_{\rm sky}}{2\pi}  \ell_{\rm max}^{2} ~. \label{eq:Fish2D_A1M_CV} 
\end{eqnarray}

\begin{widetext}

%%%%%%%%%%%%%%%%%%%%%%%%%%%%%%%%%%%%%%%%%%%%%%%%%%%%%%%%%%%%%%%%%%%%%%%%%%%%%
  \section{TripoSH decomposition of galaxy correlation functions including the wide-angle effect}
  \label{appen:Bl1l2Lx}
%%%%%%%%%%%%%%%%%%%%%%%%%%%%%%%%%%%%%%%%%%%%%%%%%%%%%%%%%%%%%%%%%%%%%%%%%%%%%

In this Appendix, by means of the TripoSH decomposition, we deal with the anisotropic signal in the galaxy correlation function without the local plane parallel approximation. Such a decomposition was already discussed in the literature \cite{Szalay:1997cc,Szapudi:2004gh,Papai:2008bd,Bertacca:2012tp,Raccanelli:2013dza}, while it was applied to the isotropic case there. We here apply it to the anisotropic case and find the signal unreported in the literature. The following equations can be derived by use of the identities in Appendix~\ref{appen:math}.

In this case, we need to treat three directions $\hat{x}_1$, $\hat{x}_2$ and $\hat{x}_{12} (\equiv \widehat{{\bf x}_1 - {\bf x}_2})$ independently. Let us introduce the TripoSH basis \cite{Varshalovich:1988ye}:
\begin{eqnarray}
  {\cal X}_{\ell \ell_1\ell_2 \ell'}^{LM}(\hat{x}_{12},\hat{x}_1,\hat{x}_2)
  &\equiv& \{ Y_{\ell}(\widehat{x}_{12}) \otimes \{ Y_{\ell_1}(\hat{x}_1) \otimes Y_{\ell_2}(\hat{x}_2) \}_{\ell'} \}_{LM} \nonumber \\
  &=& \sum_{m m_1 m_2 m' } {\cal C}_{\ell m \ell' m'}^{LM} {\cal C}_{\ell_1 m_1 \ell_2 m_2}^{\ell' m'} 
  Y_{\ell m}(\widehat{x}_{12}) Y_{\ell_1 m_1}(\hat{x}_1) Y_{\ell_2 m_2}(\hat{x}_2) ~,
\end{eqnarray}
whose orthonormality reads
\begin{eqnarray}
  \int d^2 \hat{x}_{12} \int d^2 \hat{x}_1 \int d^2 \hat{x}_2
  {\cal X}_{\ell \ell_1\ell_2 \ell'}^{LM}(\hat{x}_{12},\hat{x}_1,\hat{x}_2)
  {\cal X}_{\tilde{\ell} \tilde{\ell}_1 \tilde{\ell}_2 \tilde{\ell}'}^{\tilde{L} \tilde{M} *}(\hat{x}_{12},\hat{x}_1,\hat{x}_2)
  &=& \delta_{L, \tilde{L}} \delta_{M, \tilde{M}}
  \delta_{\ell, \tilde{\ell}} \delta_{\ell_1, \tilde{\ell}_1} \delta_{\ell_2, \tilde{\ell}_2} \delta_{\ell', \tilde{\ell}'} ~.
\end{eqnarray}
Using this, the 2-point correlator is decomposed as
\begin{eqnarray}
  \xi^s({\bf x}_{12}, \hat{x}_1, \hat{x}_2) 
  &=& \sum_{\ell\ell_1\ell_2 \ell' LM} \Xi_{\ell\ell_1\ell_2 \ell'}^{LM}(x_{12}) %% \nonumber \\
  %% &&
     {\cal X}_{\ell\ell_1\ell_2\ell'}^{LM}(\hat{x}_{12},\hat{x}_1,\hat{x}_2) ~.
\end{eqnarray}
Once the galaxy correlation function $\xi^s$ is given, the coefficients can be computed according  to
\begin{eqnarray}
  \Xi_{\ell\ell_1\ell_2\ell'}^{LM}(x_{12}) 
  &=&  \int d^2 \hat{x}_{12}  \int d^2 \hat{x}_1  \int d^2 \hat{x}_2
  \xi^s({\bf x}_{12}, \hat{x}_1, \hat{x}_2) {\cal X}_{\ell \ell_1\ell_2 \ell'}^{LM *}(\hat{x}_{12},\hat{x}_1,\hat{x}_2) ~.
\end{eqnarray}
For $\hat{x}_1 = \hat{x}_2 \equiv \hat{n}$, the TripoSH basis reduces to the BipoSH one as
\begin{eqnarray}
  {\cal X}_{\ell \ell_1\ell_2 \ell'}^{LM}(\hat{x}_{12},\hat{n},\hat{n})
= (-1)^{\ell'} 
\sqrt{\frac{(2\ell_1 + 1)(2\ell_2 + 1)}{4\pi}} H_{\ell_1 \ell_2 \ell'}
X_{\ell \ell'}^{LM}(\hat{x}_{12},\hat{n}) ~,
\end{eqnarray}
so the coefficients recover the expression under the local plane parallel approximation analyzed in Sec.~\ref{sec:formalism}, according to
\begin{eqnarray}
\xi_{\ell \ell'}^{LM}(x_{12}) 
  &=& \sum_{\ell_1\ell_2 } \Xi_{\ell\ell_1\ell_2 \ell'}^{LM}(x_{12}) 
(-1)^{\ell'} 
\sqrt{\frac{(2\ell_1 + 1)(2\ell_2 + 1)}{4\pi}} H_{\ell_1 \ell_2 \ell'} ~.
\label{eq:full2approx}
  \end{eqnarray}

We now compute the signatures from the galaxy correlation function based on Eq.~\eqref{eq:deltag_linear}, reading
\begin{eqnarray}
 \xi^s({\bf x}_{12}, \hat{x}_1, \hat{x}_2)
  &=& \int \frac{d^3 k}{(2\pi)^3} e^{i {\bf k} \cdot {\bf x}_{12}}   P_m({\bf k}) \left[b( z_1) + f(z_1) (\hat{k} \cdot \hat{x}_1)^2 \right]  %% \nonumber \\ 
%% && 
  \left[b(z_2) + f(z_2) (\hat{k} \cdot \hat{x}_2)^2 \right] ~. 
\end{eqnarray}
In the isotropic case, $P_m$ does not depend on any angles; thus, the signal vanishes except for $L = 0$, as
\begin{eqnarray}
  \Xi_{\ell\ell_1\ell_2 \ell'}^{LM}(x_{12}) &=&  i^\ell \int_0^\infty \frac{k^2 dk}{2\pi^2}
  j_\ell(kx_{12}) \Pi_{\ell \ell_1\ell_2 \ell'}^{LM}(k) ~, \\
  %------
\Pi_{\ell\ell_1\ell_2\ell'}^{LM}(k) &=&
  \sqrt{ \frac{(4\pi)^3 (2\ell+1)^2 }{(2\ell_1 + 1) (2\ell_2 + 1) } }
  c_{\ell_1} 
  c_{\ell_2} 
  H_{\ell_1 \ell_2 \ell} 
  H_{\ell \ell 0} \delta_{\ell, \ell'} \delta_{L,0} \delta_{M,0}  ~,
\end{eqnarray}
where $ c_0 = \left[b + (f / 3) \right] M_k \sqrt{P_\zeta(k)}$, $c_2 = (2/3) f M_k \sqrt{P_\zeta(k)}$ and $c_1 = c_{\ell \geq 3} = 0$. In contrast, after a bit of complicated computations, one can find that the curvature power spectrum in the quadrupolar asymmetry model \eqref{eq:zeta2_g2M} or the dipolar one \eqref{eq:zeta_A1M} creates not only $\Pi_{\ell\ell_1\ell_2\ell}^{00}$ but also
\begin{eqnarray}
\Pi_{\ell\ell_1\ell_2\ell'}^{2M}(k) &=&   
\sqrt{ \frac{(4\pi)^2 (2\ell + 1) (2\ell' + 1) }{ (2\ell_1 + 1) (2\ell_2 + 1) } }   
g_{2 M} f(k) c_{\ell_1} c_{\ell_2}  H_{\ell_1 \ell_2 \ell'} H_{\ell \ell' 2} ~,
    \end{eqnarray}
or
\begin{eqnarray}
    \Pi_{\ell\ell_1\ell_2\ell'}^{1M}(k) &=&
    \sqrt{ \frac{(4\pi)^2 (2\ell+1) (2\ell'+1)}{ (2\ell_1 + 1) (2\ell_2 + 1) } }  A_{1 M} f(k)  \nonumber \\ 
    && \sum_{J} c_{J} \left[
      c_{\ell_2} (2\ell_1 + 1)
      H_{\ell \ell_2 J }
      H_{\ell_1 J 1} (-1)^{\ell'}
      \left\{   \begin{smallmatrix}
        \ell_1 & J & 1
        \\ \ell & \ell' & \ell_2
      \end{smallmatrix}   \right\} 
      - c_{\ell_1} (2\ell_2 + 1)
      H_{\ell \ell_1 J}  
      H_{\ell_2 J 1}
      \left\{   \begin{smallmatrix}
        \ell_2 & J & 1
        \\ \ell & \ell' & \ell_1
      \end{smallmatrix}   \right\}
      \right].
\end{eqnarray}
These recover the real-space expressions of Eqs.~\eqref{eq:PlldLM_iso}, \eqref{eq:PlldLM_g2M} and \eqref{eq:PlldLM_A1M} via Eq.~\eqref{eq:full2approx}.

\end{widetext}

%%%%%%%%%%%%%%%%%%%%%%%%%%%%%%%%%%%%%%%%%%%%%%%%%%%%%%%%%%%%%%%%%%%%%%%%%%%%%
\section{Useful identities}\label{appen:math}
%%%%%%%%%%%%%%%%%%%%%%%%%%%%%%%%%%%%%%%%%%%%%%%%%%%%%%%%%%%%%%%%%%%%%%%%%%%%%

We here summarize the identities used for the derivations of the equations.

The angular dependences in functions can be decomposed into spherical harmonics according to, e.g., 
\begin{eqnarray}
  {\cal L}_l(\hat{k} \cdot \hat{n}) &=& \frac{4\pi}{2l+1} \sum_m  Y_{lm}(\hat{k}) Y_{lm}^*(\hat{n}), \label{eq:L_expand}
\end{eqnarray}
and 
\begin{eqnarray}
  e^{i{\bf k} \cdot {\bf x}} &=& \sum_{L M}
  4\pi i^{L}   j_L(kx) Y_{LM}(\hat{k}) Y_{LM}^*(\hat{x}) ~. \label{eq:e_expand}
\end{eqnarray}

Via the addition rule of spherical harmonics
\begin{equation}
  Y_{l_1 m_1}(\hat{n}) Y_{l_2 m_2}(\hat{n}) = \sum_{l_3 m_3} Y_{l_3 m_3}^*(\hat{n}) h_{l_1 l_2 l_3} \left(
  \begin{smallmatrix}
  l_1 & l_2 & l_3 \\
  m_1 & m_2 & m_3 
  \end{smallmatrix}
 \right), \label{eq:Ylm_add}
\end{equation}
and the orthonormal condition
\begin{eqnarray}
  \int d^2 \hat{n} Y_{l_1 m_1}(\hat{n}) Y_{l_2 m_2}^*(\hat{n}) = \delta_{l_1, l_2} \delta_{m_1, m_2} ~, \label{eq:Ylm_ortho}
  \end{eqnarray}
one can integrate the product of any number of spherical harmonics.

The angular momenta in the Wigner symbols arising from the angular integrals of spherical harmonics can be added as
\begin{equation}
\sum_{l_3 m_3} (2l_3+1)
  \left(
  \begin{smallmatrix}
  l_1 & l_2 & l_3 \\
  m_1 & m_2 & m_3
  \end{smallmatrix}
  \right)
  \left(
  \begin{smallmatrix}
  l_1 & l_2 & l_3 \\
  m_1' & m_2' & m_3 
  \end{smallmatrix}
  \right) = 
\delta_{m_1, m_1'} \delta_{m_2, m_2'} , \label{eq:3j_sum_l3m3}
\end{equation}
%------
  \begin{eqnarray}
 \sum_{m_1 m_2}
  \left(
  \begin{smallmatrix}
  l_1 & l_2 & l_3 \\
  m_1 & m_2 & m_3
  \end{smallmatrix}
  \right)
  \left(
  \begin{smallmatrix}
  l_1 & l_2 & l_3' \\
  m_1 & m_2 & m_3' 
  \end{smallmatrix}
  \right)
  = \frac{\delta_{l_3, l_3'} \delta_{m_3, m_3'}}{2l_3+1}, \label{eq:3j_sum_m1m2}
  \end{eqnarray}
  %----
  \begin{eqnarray}
  \sum_{m} (-1)^{l - m}
  \left(   \begin{smallmatrix} l & l & L \\ m & -m & 0\end{smallmatrix}   \right) &=& \sqrt{2l + 1}\delta_{L,0} ~, \label{eq:3j_sum_m}
\end{eqnarray}
  and
  %----
\begin{eqnarray}
&& \sum_{m_4 m_5 m_6} (-1)^{\sum_{i=4}^6( l_i - m_i) }
\left(
  \begin{smallmatrix}
  l_5 & l_1 & l_6 \\
  m_5 & -m_1 & -m_6 
  \end{smallmatrix}
 \right) 
\left(
  \begin{smallmatrix}
  l_6 & l_2 & l_4 \\
  m_6 & -m_2 & -m_4 
  \end{smallmatrix}
  \right)
  \nonumber \\
&&\qquad \times
\left(
  \begin{smallmatrix}
  l_4 & l_3 & l_5 \\
  m_4 & -m_3 & -m_5 
  \end{smallmatrix}
 \right)%%  \nonumber \\
 %% &&\quad
 = \left(
  \begin{smallmatrix}
  l_1 & l_2 & l_3 \\
  m_1 & m_2 & m_3 
  \end{smallmatrix}
 \right) 
\left\{
  \begin{smallmatrix}
  l_1 & l_2 & l_3 \\
  l_4 & l_5 & l_6 
  \end{smallmatrix}
 \right\} . \label{eq:3jto6j}
 \end{eqnarray}

%############################################################################
% Create the reference section using BibTeX
\bibliography{paper}

\end{document}